\documentclass[letterpaper, nonacm]{acmart}

\AtBeginDocument{%
  \providecommand\BibTeX{{%
    \normalfont B\kern-0.5em{\scshape i\kern-0.25em b}\kern-0.8em\TeX}}}

\usepackage{hyperref}
\usepackage{breakurl}

\begin{document}

\title[Legitimate Power, Illegitimate Automation]{Legitimate Power, Illegitimate Automation:}
\subtitle{The problem of ignoring legitimacy in automated decision systems}

\author{Jake Stone}
\email{Jake.stone@anu.edu.au}
\affiliation{%
  \institution{The Australian National University}
  \country{Australia}
}

\author{Brent Mittelstadt}
\email{brent.mittelstadt@oii.ox.ac.uk}
\affiliation{%
  \institution{The Oxford Internet Institute}
  \country{United Kingdom}
}

\authorsaddresses{}

\begin{abstract}
 Progress in machine learning and artificial intelligence has spurred the widespread adoption of automated decision systems (ADS). An extensive literature explores what conditions must be met for these systems' decisions to be fair. However, questions of legitimacy --- why those in control of ADS are entitled to make such decisions --- have received comparatively little attention. This paper shows that when such questions are raised theorists often incorrectly conflate legitimacy with either public acceptance or other substantive values such as fairness, accuracy, expertise or efficiency. In search of better theories, we conduct a critical analysis of the philosophical literature on the legitimacy of the state, focusing on consent,  public reason, and democratic authorisation. This analysis reveals that the prevailing understanding of legitimacy in analytical political philosophy is also ill-suited to the task of establishing whether and when ADS are legitimate. The paper thus clarifies expectations for theories of ADS legitimacy and charts a path for a future research programme on the topic.
\end{abstract}

\maketitle

\section{Introduction}
The central question in the study of legitimacy is: in a society of free and equal individuals, why are some permitted to hold power over others? Theories of legitimacy seek to answer this question by specifying the normative conditions which must be satisfied to reconcile grants of power with the freedom and equality of those under its influence, thereby rendering power morally permissible \cite{Viehoff2019, greene2016, RawlsPL, LazarCBC, Pettit2012, Buchanan2002}.  

Philosophers who explore this topic have long concentrated on the state. Yet, beyond the state ADS operated by private individuals have gained significant power over our lives. They are used to make decisions that affect areas such as finances, career paths, and access to healthcare and education. Given this influence, it’s prudent to ask: why are the those with control over these systems permitted to wield such power?

Unfortunately, questions of ADS legitimacy have often failed to receive the attention they deserve. Pioneering work in the 2010s highlighted how these systems often favoured certain social groups, creating or reinforcing inequitable relationships and distributions of resources \cite{Angwin2016, eubanks2018, Oneil2016, Lum2016, Gebru2018}. While this focus on systems and their outcomes was warranted, it had the unfortunate effect of pushing questions about their underlying legitimacy to the periphery. This is evident in much of the fair machine learning literature where scholars propose statistical criteria as a means of making the exercise of power fair without questioning whether the power should be conferred in the first place \cite{Barocas2023}. 

It should be noted that legitimacy has not been entirely ignored in the literature to date. Critical analyses of ADS, especially those from scholars working in feminist and afro-feminist traditions, often include calls for the abolition of certain types of ADS  or guarantees that power is vested only in those entitled to it \cite{Benjamin2019, Klein2023, Birhane2021, Noble2018, hampton2021}. Such calls are in spirit, if not in name, based on legitimacy. Additionally, a growing body of work uses the concept of legitimacy explicitly when critiquing ADS \cite{CaloandCitron2020, Wang2022, Boyd2020, Purves2022, Waldman2022, Waldman2019, Barocas2023}.\footnote{For another source that discusses legitimacy in the context of technology generally, but not specifically as it applies to ADS, see Taylor \citeyearpar{taylor2021}. Note that this paper conflates the evaluative criteria of power, an issue for which we criticise the ADS legitimacy literature in Section 3.} Nonetheless, legitimacy has largely been invoked in sociotechnical critiques of ADS without engaging with the first-order literature on legitimacy in political philosophy, or in the normative philosophy of computing. As a result, the literature frequently relies on underdeveloped notions and as such does not provide us with the conceptual tools needed to assess how ADS power should be legitimated.  

This gap demands immediate remedy because legitimacy determines whether a system should have power over individuals, while fairness examines how this power should be exercised. That is, questions of legitimacy are prior to questions of fairness. Therefore, as Barocas, Hardt, and Narayanan \citeyearpar{Barocas2023} observe, discussing fairness without first establishing legitimacy is a tacit endorsement of a decision-maker’s power and as such reinforces illegitimate power holders.\footnote{Keyes, Hutson, and Durbin \citeyearpar{keyes2019} make a similar point.}  This lack of scrutiny unintentionally cedes ground to private entities which are increasingly encroaching on the role of the state and impacting upon citizens’ rights. The recent trial of GOV.UK Chat --- a chatbot powered by ChatGPT to assist the public in answering queries on UK government websites --- is a case in point \cite{GDS2023}.  While the chatbot may improve efficiency and user convenience, entrusting a private entity with control over the public's ability to engage with government services grants OpenAI concerning and perhaps illegitimate power. Consequently, a thorough legitimacy assessment should precede any discourse on potential benefits. 

\subsection{Assumptions, definitions, and scope}
Before proceeding, a brief aside is needed on the conceptualisation and scope of legitimacy in this paper. ADS include all sociotechnical systems that utilise machine learning or other computational methods to automate, either wholly or in part, the process of making decisions about people in public or private settings. Examples include university admissions, pre-trial detention, predictive policing, employee screening, immigration, and healthcare recommendations. 

Legitimacy, as we conceptualise it, is a two-place relation that specifies the conditions of morally permissible power between two parties: the decision-makers who wield power via ADS and the decision subjects to whom that power is applied. Power, broadly understood, is the ability to unilaterally make decisions which affect the interests of others \cite{LazarCBC, LazarVoE}. Such power is exercised by the end-users who make decisions using ADS \textit{and} the technical specialists and firms who make decisions about the design and deployment of these systems. Consequently, the theory presented here applies to any agent who has the capacity to make decisions about the design, deployment, or use of ADS which substantively affect the interests of decision subjects. This means that for any single system there will be multiple agents who are subject to the requirements of legitimacy. 

Given the substantial difference between the power of ADS decision-makers and that of the state, one might doubt that such power requires legitimation. Yet, many of the same considerations that make us care about legitimating the state also apply in the case of ADS. Power, whether exercised by the state or private entities, shapes the lives of those it governs. In the context of ADS, power affects our career opportunities, educational, financial, welfare, and healthcare resources, and, notably in pretrial risk assessments, our right to liberty.

This is not meant to imply that ADS must be subject to the same legitimisation methods as the state. We adopt a pluralist account of legitimacy, in which the appropriate legitimation method depends on the type of power in question and the costs of various legitimation processes \cite{LazarCBC, LazarVoE, Himmelreich2023}. Considering the diversity of ADS use cases it seems likely that consent will legitimate in some cases, public reason in others, democratic authorisation in others still. Hence, we do not set out to identify a single correct legitimation method, but to investigate different theories and the obstacles they encounter when applied to ADS.

Nor do we argue that all ADS exercising power are illegitimate or that new legitimation processes are necessarily required. People regularly use power that has previously been legitimised. For example, companies have the ability to hire and fire employees in accordance with laws established and legitimated by a national government. In our analysis we assume such existing legitimate powers do not require additional legitimation. This means that in many cases ADS decision-makers --- for example those pricing insurance premiums, determining the conditions of a loan, or screening job applicants --- may already be legitimised.

Consequently, the target of our analysis is new power relations which are facilitated by the design, deployment, and use of ADS. Such power relations often arise because automating decision processes requires external experts. These third-parties make substantive decisions about policy \cite{Barocas2023, Huq2020, Citron2007, Waldman2019, Passi2019, Eaglin2016} but, unlike the original institutions responsible for enacting policy, they have not undergone legitimation. Furthermore, existing decision-makers may find their power changed or expanded through ADS, enabling different decisions or substantially different decision-making processes. Whether there is a change in the parties to a relation, or a change in the power they wield, a new relationship is formed which requires legitimation.\footnote{This argument contradicts Himmelreich \citeyearpar[p.1339]{Himmelreich2023} who claims that AI does not give rise to legitimation requirements because it is used in already coercive systems or does not increase the pervasiveness of such coercion. This objection misunderstands the problem of legitimacy. The issue isn’t one of coerciveness or pervasiveness but that we have a new relation, either because the parties to the relation or the kind of power over a decision subject has changed, and this relation requires legitimation.}  

Consider, for example, the COMPAS pretrial detention algorithm \cite{Angwin2016}. Northpointe was given the power to influence pretrial detention decisions, in effect transferring power to them which was previously held by the state. COMPAS likewise changed the nature of judicial decision-making by allowing judges to rely on algorithmically generated risk assessments, thereby introducing new complexity and opacity to the exercise of judicial power.

Although this paper explores ways to legitimate systems, it is important to acknowledge that the presence of such methods does not obligate us to legitimate ADS. In many use cases abolition, not legitimation, will be more appropriate. We intend for this paper to aid in deciding which path to pursue; the theories examined here can strengthen arguments for abolition by demonstrating when ADS are incapable of satisfying the legitimating criteria.

With regard to scope, the political philosophy of legitimacy is a voluminous area of scholarship. Consequently, this paper does not, nor could it, survey all of the theories and debates present in the philosophical literature. Instead, we focus on theories that are either particularly intuitively promising or which have already received some attention: (1) consent, (2) public reason, and (3) democratic authorisation. Our hope is that in advocating for the importance of ADS legitimacy and providing a primer on the topic we will have laid the groundwork for future research on these, and other, theories. 

\subsection{Structure of the paper}
We begin by clarifying the target of our inquiry in sections 2 and 3. Section 2 argues that scholars of technology have adopted the sociological concept of descriptive legitimacy  which leads them to ascribe legitimacy based on public beliefs (which can be mistaken) about compliance with the conditions for morally justified power.\footnote{See for example: boyd \citeyearpar{Boyd2020}; Ferreira, Merendino, and Meadows \citeyearpar{Ferreira2021}; Waldman and Martin \citeyearpar{Waldman2022}.} Moreover, the unique features of ADS increase the likelihood of such mistakes occurring. Focusing on descriptive legitimacy can thus grant decision-makers a veneer of legitimacy without the need to actually fulfil the essential criteria for morally permissible power. 

Section 3 argues that confusion also results from a conflation of legitimacy with other normative considerations such as fairness, accuracy, expertise, and efficiency.\footnote{Examples include: Waldman \citeyearpar{Waldman2019}; Calo and Citron \citeyearpar{CaloandCitron2020}; Wang et al. \citeyearpar{Wang2022}; Barocas, Hardt and Narayanan \citeyearpar{Barocas2023}.} By focusing on them theorists inadvertently reinforce illegitimate power. Instead, we argue that discussion should be reoriented to focus on the fundamental question: what criteria reconcile the conflict of ADS decision-making power with the freedom and equality of decision subjects?

Having cleared the way to address this fundamental question, the analysis then turns to different theories of normative legitimacy and their application to ADS in sections 4 through 6. In Section 4 we find that ADS, due to their opacity and involuntariness, often fall short of a consent-based standard. Section 5 argues that public reason theories have a role to play but are unlikely to provide answers to many of the questions encountered in legitimating ADS. On a more positive note, Section 6 presents a cautiously optimistic analysis of theories of ADS legitimation grounded in democratic authorisation. 

\section{Descriptive and normative legitimacy }
To be legitimate a power relationship must fulfill the requisite conditions of moral permissibility, which we refer to as the “normative criteria.” This is not meant to imply any kind of moral realism, that the normative criteria are socially constructed is perfectly consistent with this claim. Our point is that that there are normative conditions, however they are formed, which an entity must satisfy in order for their possession of power to be legitimate. This section explains how descriptive theories of legitimacy concentrate on public beliefs and thus overlook the need to satisfy normative criteria. Furthermore, because of the features of ADS it is particularly likely that descriptive assessments of them will diverge from our normative assessments. 

To be clear, descriptive theories should not be abandoned. In fact, such theories are important tools for challenging inequitable distributions of power (see: Section 2.3). Rather, work on accountability in ADS needs to broaden its focus and allocate more time to first-order theorising about, and assessments of, normative legitimacy.

\subsection{What is descriptive legitimacy?}
Perhaps the most influential account of descriptive legitimacy can be found in the work of Max Weber \citeyearpar{Weber1946, Weber1947, Weber2019} for whom “the basis of every system of authority, and correspondingly of every kind of willingness to obey, is a belief, a belief by virtue of which persons exercising authority are lent prestige” \cite[p.382]{Weber1947}. According to such theories an institution is \textit{descriptively} legitimate to the extent that the public believes it to be so.  

This definition, or something similar, has been widely adopted by social scientists \cite{Applbaum2004, Simmons1999, Purves2022, Suchman1995}. Applbaum \citeyearpar{Applbaum2004} criticizes this trend, arguing that Weber was aware of his focus on the descriptive dimensions of legitimacy but that later readings have blurred this distinction, failing to adequately separate the normative aspect (the permissibility of power) from the descriptive aspect (beliefs about such permissibility).

Scholars working on sociotechnical systems have largely continued in this vein. Consider, for example, boyd \citeyearpar[p.260]{Boyd2020} who writes: “Data’s legitimacy comes from a belief that we can collectively believe that those data are sound, valid, and fit for use.” Or Ferreira, Merendino, and Meadows \citeyearpar[p.1083]{Ferreira2021}: “Legitimacy consists of a judgement about the desirability, propriety or appropriateness of the actions of an entity --- such as a company --- vis-\`a-vis prevalent values and meanings in society.” Or Waldman and Martin \citeyearpar[p.1]{Waldman2022} who advocate for a greater focus on the descriptive when they write: “The algorithmic accountability literature … elides a fundamentally empirical question: whether and under what circumstances, if any, is the use of algorithmic systems to make public policy decisions perceived as legitimate?”

These are descriptive conceptions of legitimacy. But to be fair, it is often unclear whether the authors are (1) merely focusing on public beliefs to the exclusion of normative criteria,\footnote{Something which we can have good reason to do, see Section 2.3.}  (2) conflating the normative with the descriptive, or (3) suggesting legitimacy is solely determined by beliefs.\footnote{For an excellent exception to this lack of clarity see Purves and Davis \citeyearpar{Purves2022}.}  Regardless of intent, this tendency focuses analysis on empirical questions to be answered not by whether the relevant norms have been satisfied but by whether the public believes they have been \cite{Applbaum2004}.  This manifests in the literature in a preoccupation with the determinants of public beliefs towards institutions and how those beliefs affect institutional behaviour \cite{Purves2022}.\footnote{For other examples of this empirical focus see: Danaher et al. \citeyearpar{Danaher2017}; de Fine Licht and de Fine Licht \citeyearpar{De2020}; Larsen \citeyearpar{Larsen2021};  Saeed, Riaz, and Baloch \citeyearpar{Saeed2022}.}

\subsection{The problem with descriptive legitimacy}
The problem with this preoccupation is that legitimacy debates are not, at least centrally, about what people believe but rather about whether a decision-maker actually satisfies the normative criteria which justify their power \cite{Buchanan2006, Applbaum2004, Simmons1999}.  When someone claims, for example, “the use of facial recognition to identify suspects of crimes is illegitimate,” they are not making a claim about whether others believe such practices are illegitimate; rather, they are claiming that such uses \textit{are} illegitimate. 

Normative and descriptive assessments of legitimacy overlap in practice. Beliefs about a decision-maker’s legitimacy often track whether they satisfy relevant normative criteria. But this means that the normative criteria are, in some sense, prior to public beliefs \cite{Applbaum2004, Simmons1999}. Consequently, examining only whether a decision-maker has public support fails to answer the prior question of precisely which criteria the public believe have been satisfied \cite{Buchanan2006, Buchanan2002} and misses an opportunity to scrutinize the principles that inform legitimation \cite{Thomas2014}.  

It could be argued that this distinction is practically irrelevant; if public beliefs accurately track satisfaction of the criteria, assessing normative legitimacy should be as simple as gauging a decision-maker’s degree of public support. In such cases legitimacy can be determined without knowing the content of the normative criteria used. Nothing of importance would appear to turn on our knowledge of them. Unfortunately, numerous cases exist in which public beliefs and the satisfaction of normative criteria diverge. Take for example Volkswagen,  Wells Fargo,  or Enron,  which were all regarded as upstanding corporate citizens before their misconduct became public knowledge \cite{Hotten2015, Mclean2017, Petroff2015, Fox2003, Thomas2002}. 

This divergence between public beliefs and normative reality is particularly acute in the case of ADS. One key reason is what we will refer to as ‘embeddedness’: the more an institution integrates into our daily lives, the more reluctant we are to re-evaluate its legitimacy \cite{Culpepper2020}.  Ongoing interactions with an institution normalise it over time, leading to reduced negative legitimacy evaluations and a tendency to overlook unethical behaviour \cite{Tost2011}.  Given the ubiquity of ADS and their importance in delivering many vital services, decision-makers utilising ADS stand to benefit from this phenomenon. This may be exacerbated when ADS are part of digital platforms that have themselves become essential to daily life \cite{Culpepper2020}, effectively granting dual layers of legitimacy protection through embeddedness. 

Apart from embeddedness, other factors also cast doubt on the public's ability to accurately gauge the legitimacy of ADS decision-makers. For example, although much research tries to counter this perception, the use of ADS conveys an illusion of objectivity, thereby implying moral permissibility \cite{Barocas2023}. Similarly, the use of big data, a fundamental requirement for training ADS, can elevate perceived legitimacy \cite{Ferreira2021}. Lastly, while states must meet transparency requirements that allow at least some public scrutiny, ADS are often proprietary \cite{Citron2014}  which makes informed public assessment difficult.

In some cases public beliefs will accurately track the normative status of a decision-maker. In such cases descriptive legitimacy can be used as a rough heuristic for normative legitimacy. However, the issues raised above make it imprudent to rely solely on public beliefs when assessing legitimacy. ADS-specific normative legitimacy criteria should instead be constructed independently from prevailing public attitudes and used to audit decision-makers for compliance with these standards.

\subsection{In defence of descriptive legitimacy}
If descriptive legitimacy is only a rough heuristic for the thing we actually care about (i.e., normative legitimacy), why continue to use it? One reason is clear: it is instrumentally valuable. The willingness of the public to support and interact with ADS is contingent upon their perception of decision-makers’ legitimacy \cite{Ferreira2021, Purves2022, Tost2011}. Consequently, understanding the conditions of public beliefs and effective legitimacy management are vital to enable long-term use of ADS \cite{Wang2022, Barocas2023, CaloandCitron2020}. 

On this account descriptive legitimacy may appear to amount to little more than an exercise in public relations. However, given that the ability to continue operating depends on public support, measures of descriptive legitimacy can pressure decision-makers to adopt structures and behaviours that are seen as legitimate \cite{Dacin1997, Dimaggio1983, Tolbert1983, Meyer1977}. This also makes descriptive legitimacy crucial for activists seeking to dismantle existing power structures and abolish particular uses of ADS. In this way, descriptive legitimacy can help guide behaviours and bring about important social change \cite{Meyer1977}. Take, for example, campaigns against the use of sweatshop labour by Nike and Apple \cite{Porter2012, Bartley2011}. 

Nevertheless, descriptive theories leave the fundamental question of what grounds legitimacy unanswered. Potential candidates can be found in literature on the legitimacy of ADS which focus on fairness, accuracy, expertise, and efficiency. In the following section we explain why these criteria should not be confused with those that legitimate ADS. 

\section{The evaluative criteria of power}
Power is subject to different evaluative criteria and it is important to distinguish between them when addressing issues of legitimacy \cite{Simmons1999, Buchanan2002, Adams2018}. Recent work by Seth Lazar provides a useful categorisation of the kinds of evaluative questions philosophers have posed about power: we might ask whether it is exercised (1) by the appropriate person; (2) in the correct way; and (3) for the right ends \cite{LazarCBC, LazarVoE}. Following Lazar we will refer to these as the ‘who’, ‘how’, and ‘what’ questions. 

When we ask ‘who’ questions, our task is to identify the normative ground of a given decision-maker’s power. We might ask, for example: why is it Magistrate Smith, and not Joe, who is entitled to decide my bail conditions? These are questions of legitimacy as authorisation. On the other hand, ‘how’ questions are concerned not with the entitlement to power but with the manner in which it is exercised. They prescribe procedural and substantive conditions which must be met before, during, and after the exercise of power such as notice and opportunity for comment, consistency and transparency, and avenues for contestability. ‘What’ questions restrict the objectives for which power holders, once legitimate, can utilise their power. Acceptable objectives are those which are consistent with accepted moral and societal norms \cite{LazarCBC, LazarVoE}. 

\subsection{Independence of the evaluative criteria}
According to some theorists power's legitimacy is determined by its ability to achieve right ends in a manner that cannot be improved upon by feasible alternatives \cite{Raz1986, Raz2005, Wellman1996, Arneson2003}. However, many philosophers find such views unsatisfactory, arguing that answers to ‘how’ or ‘what’ questions do not provide satisfactory answers to ‘who’ questions \cite{Simmons1999, Kolodny2014II, Kolodny2016, LazarCBC, LazarVoE, Adams2018}. We endorse this latter position and contend that focusing solely on the manner of power's exercise or the ends it achieves leaves behind a residual moral complaint, which must be resolved by other means. 

For an intuitive understanding of this point, consider the following example: suppose you are exceptionally well-placed to make decisions about your friend’s financial life. You have an excellent track record of investing, care about their wellbeing and, if given the chance, would make such decisions in a procedurally and substantively correct way. Your friend, on the other hand, is drawn to high-risk investments and spends impulsively, often overlooking options that present better long-term financial stability and growth. Under these circumstances, you would make better choices regarding your friend’s finances than they do. However, this capability in no way entitles you to manage their financial affairs. This observation is true in many areas of life: even if you have an instinct for strategic moves in a chess game, a knack for picking the fastest queue in a supermarket, or a talent for planning efficient travel itineraries, these capabilities alone do not entitle you to make those decisions for others.

These examples reflect a key difference between characteristics of power and the reasons someone is entitled to wield such power. ‘How’ constraints require power to be exercised in specific ways. Similarly, ‘what’ constraints dictate that power should only be exercised in pursuit of acceptable goals. Therefore, the ‘how’ and ‘what’ questions determine the characteristics of power. Legitimation processes do not directly address the characteristics of a power, but rather why certain decision-makers are or should be entitled to wield power with such characteristics. 

This observation does not mean that exercising power in the right way or pursuing admirable goals have no bearing on the justification of power. ‘How’ and ‘what’ constraints can make the exercise of illegitimate power, all things considered, permissible.  Take, for example, laws against murder in apartheid South Africa. Despite the regime’s clear illegitimacy, there is an overriding ‘what’ reason (the prevention of murder) for allowing the state to enforce this law. That this reason justifies the state’s use of power, does not address the issue of its illegitimacy, which still demands resolution \cite{Simmons1999, Kolodny2014II, Kolodny2016, LazarCBC, LazarVoE}.  

Power can be restricted so that it must be exercised correctly and in pursuit of proper goals. In such cases, unfair or ill-motivated exercises would be \textit{ultra vires} and therefore illegitimate. In this way ‘how’ and ‘why’ can serve to delegitimate. But constraints on the exercise of power usually allow for some latitude, acknowledging human fallibility and the complexities of decision-making \cite{Wellman2023}. Further, in legitimation processes not involving the state, fairness or acceptable goals may not be required at all: the birthday girl can choose whatever she wants for dinner regardless of her brother’s preferences! 

\subsection{Conflation in the literature}
This point is worth emphasising because recent scholarship on the legitimacy of ADS focuses on how and for what purpose power is exercised while leaving ‘who’ questions largely unaddressed. For example, Barocas, Hardt, and Narayanan \citeyearpar{Barocas2023} argue that several features of automated decision making systems limit predictive accuracy and as such have the ability to undermine legitimacy. Similarly, Wang et al. \citeyearpar{Wang2022} assert that machine learning's inherent traits negate efficiency, accuracy, and fairness claims made by ADS developers thus rendering them illegitimate. Waldman \citeyearpar{Waldman2019} argues against efficiency but identifies fairness as one of the key requirements of legitimacy for ADS. Lastly, Calo and Citron \citeyearpar{CaloandCitron2020} argue that ADS undercut the bases --- efficiency and expertise --- for the delegation of administrative power. They contend that delegates often lack a nuanced understanding of these systems, which weakens their claims to expertise, while the high incidence of errors erodes the argument for efficiency. 

Efficiency, accuracy, and fairness are straightforwardly ‘how’ constraints as they specify minimum conditions which must be met before, during, and after the exercise of power. Efficiency requires that decision-makers minimise the time and resources used to make decisions. Fairness, either procedural or substantive, requires that the way in which decisions are made or the distribution of outcomes from those decisions are equitable \cite{Wachter2020}. Accuracy requires that decisions correspond to the reality they are intended to represent. 

Expertise, although not instantly recognisable as a ‘how’ or ‘what’ constraint, operates as both in practice. Calo and Citron’s description of expertise is best understood as a ‘how’ constraint. According to their theory the legitimacy of administrative bodies is undermined by algorithmic systems when administrators lack the necessary expertise to understand how those systems function, leading to erroneous decisions and an inability to provide explanations sufficient for due process. That is, administrators lacking expertise cannot ensure that the ‘how’ – the manner and quality of decision-making – is adequate. Expertise is also closely linked to ‘what’ as it helps ensure that power is exercised in pursuit of acceptable goals; we delegate decision-making capacity to experts because they understand our desired policy objectives, thus enabling them to align decisions with these goals. Barocas, Hardt and Narayanan \citeyearpar{Barocas2023} don’t explicitly deploy expertise as a ground of legitimacy. However, they do have related discussion which argues that to be legitimate decision-makers must be able to justify their predictive targets in terms of acceptable goals, framing part of legitimacy as a ‘what’ constraint. And, as the preceding discussion makes clear, how and what do not legitimate. 

Focusing too narrowly on ‘how’ and ‘what’ questions should be avoided because it suggests that if decision-makers can utilise ADS over which they have sufficient expertise, and the systems and decision-making procedures are fair, efficient, and accurate they would be legitimate, thereby granting them underserved power. To be fair, Wang et al. \citeyearpar{Wang2022} argue that accuracy, efficiency, and fairness are necessary but not sufficient conditions of legitimacy, meaning that uses of ADS that meet these criteria are not automatically legitimate. Nevertheless, emphasising these factors may inadvertently reinforce illegitimate power by focusing on criteria which do not legitimate. Instead, the legitimation of ADS should be grounded in the satisfaction of criteria which reconcile power with the freedom and equality of individuals, i.e., answers to ‘what’ questions. 

In the next section we commence the search for such criteria with theories of legitimacy grounded in consent. This is advisable because many of the decision-makers who currently wield the greatest degree of power through ADS are private actors. In the private domain consent is often used to establish or modify relations of power, for example via contract. As will come as no surprise to those familiar with the application of consent in other domains,\footnote{See for example: Barocas and Nissenbaum \citeyearpar{Barocas2014}; Lanier \citeyearpar{lanier2014}; Sadowski \citeyearpar{sadowski2019}.} realising valid consent in practice for ADS is challenging. 

\section{Consent}
The argument for consent as a ground of legitimacy proceeds as follows. Each citizen should be free to dictate the course of their own lives \cite{Simmons1976, Simmons1979, Raz1987}.  However, the state wields power to regulate social conduct, maintain order, and enforce laws. This poses a challenge to individual freedom; if I happen to disagree with a law or regulation, then I will be forced by the state into compliance, thus infringing my freedom. If, however, I choose to surrender some of my freedom to the state, recognising that this grant serves my best interests, then the state's power becomes an extension of my freedom rather than its suppression, thus rendering it legitimate. 

\subsection{Applying consent}
For consent to be effective it must be voluntary and informed \cite{Nuremberg1949}. To be informed someone must have sufficient knowledge of the proposed course of action to arrive at a well-considered decision \cite{Nuremberg1949}. To be voluntary, the person granting consent must have reasonable alternatives \cite{Simmons1979, Wertheimer1990}. ADS pose a problem for both elements: the opacity of ADS undermines well-informed consent, and the absence of reasonable alternatives in many use cases compromises voluntariness.

With regards to opacity, ADS are operationally opaque because they are proprietary in nature, and it is often unclear to affected individuals when or how they are deployed. In such cases consent is absent as by definition a person cannot consent to something of which they are unaware. Beyond operational opacity, many ADS are also technically opaque. While many ‘explainability’ methods have been developed in recent years that make machine learning and AI models more directly interpretable or explain post-hoc how they produce outputs from inputs, their utility for non-expert end-users in different ADS use cases remains an open question requiring further study \cite{Burrell2016, Rudin2019}. There may likewise be inherent limitations on human understanding of complex models (which can be composed of millions of interdependent features), meaning explainability methods would at best provide an incomplete or highly simplified global understanding of ADS functionality, or be limited to explaining specific granular relationships or individual outputs \cite{Mittelstadt2019explaining, Wachter2017}. 

Moreover, the things which need to be explained to meet the informed criterion are not limited to technical details. Being informed arguably also requires information about the data used to train a model, how the model was incorporated into other sociotechnical infrastructures, and the consequences of these choices for the likely outcomes a decision subject will receive \cite{Gebru2021, Mitchell2019}. Overcoming opacity in ADS to achieve informed consent is therefore not simply a problem of developing more interpretable model architectures or better post-hoc explanation methods, but also requires consistent and informative documentation and disclosures to affected parties \cite{Mittelstadt2021interpretability}. 

Even if these opacity issues can be overcome, a more serious challenge to consent is that ADS are often involuntarily imposed on decision subjects. This can occur in multiple ways. First, the goods or services a decision subject is seeking may only be available from one provider who uses ADS. For instance, when a defendant is subject to a risk assessment tool used in the criminal justice system. Secondly, if all of the providers in a particular sector utilise ADS then there is no way to access those services without being subject to some form of ADS. Lastly, while some sectors might have providers who do not use ADS, it is often impractical or infeasible to access them \cite{Daniels1997}. For instance, finding a public school without algorithmic admissions processes might be possible in theory, but the distance to that school may be impractical. 

This last source of involuntariness is particularly important because those who are socially or economically marginalised will have fewer reasonable alternatives available to them and as such will more likely be involuntarily subject to ADS. Consequently, if we ground the legitimacy of ADS in consent the economically and socially marginalised are more likely to be subjected to illegitimate systems. 

In their current form many ADS would be illegitimate according to a consent-based standard. But are these in-principle or merely contingent objections? Some of these problems seem to be solvable, for example, by promoting better practices around operational opacity. However, concerns about technical opacity seem less tractable. Even if technical methods to explain the outputs of ADS improve substantially there is likely to remain a portion of the public who cannot or will not understand how ADS operate to a degree sufficient for meaningful consent \cite{Gillespie2014}.  

As concerns involuntariness all of the cases discussed above involve situations where expecting decision subjects to forego the goods or services at hand is unreasonable. However, when the goods or services are less important, choosing to forego them may be a reasonable alternative, thereby increasing the potential to gain valid consent. At best, consent is a viable way to legitimate low-risk or less consequential ADS, but remains inappropriate in higher-stakes situations.  

Following the preceding discussion, two options remain to establish ADS legitimacy. The first is to view ADS legitimacy as infeasible in many high-stakes use cases. The second is to find other ways to legitimate these systems. Taking the second option means looking for approaches beyond consent to reconcile the conflict between power and individual freedom and equality. In pursuit of such a strategy, we discuss a ‘public reason’ framework next which is viewed as a promising response to the limitations of consent and pathway to make algorithmic decisions legitimately \cite{Binns2018}. 

\section{Public Reason}
For John Rawls \citeyearpar{RawlsPL} a need for public reasons as a criterion of legitimacy follows from the assumption that citizens are reasonable and accept the burdens of judgment. Reasonable individuals only propose rules for collective governance which they themselves would endorse \cite[p.54]{RawlsPL}. Accepting the burdens of judgment requires one to acknowledge that reasonable and rational people will disagree on important moral and political issues due to the complexity of these questions and the diversity of their experiences and perspectives \cite[pp.55--7]{RawlsPL}.  From these two assumptions follows the principle of legitimacy: power is legitimate only if it is exercised in accordance with a constitution the essentials of which no reasonable person would reject \cite[p.217]{RawlsPL}.  

This necessitates the ability to justify the essentials of a constitution using public reasons, which “appeal only to presently accepted general beliefs and forms of reasoning found in common sense, and the methods and conclusions of science when these are not controversial” \cite[p.224]{RawlsPL}.  Examples include the equality of citizens in basic political rights and liberties, freedom of association and speech, and the rule of law. In contrast, private reasons are rooted in comprehensive doctrines of individuals comprised of moral, religious, and philosophical values which can vary among reasonable people and may thus be rejected.

\subsection{Applying public reason}
Binns \citeyearpar{Binns2018} argues that in situations where decision-makers and decision subjects disagree public reasons offer a common ground for resolution and in so doing can act as a constraint on legitimate decision-making. 

The problem with legitimating ADS in this way is that it overextends public reason by requiring it to resolve disagreement on a broad range of moral and political questions for which it is poorly suited. Binns touches upon the issue of extension, suggesting that public reason might be “a constraint on the legitimate exercise of decision-making power in general” \cite[p.549]{Binns2018}. However, his brief discussion belies considerable uncertainty about whether such extension is feasible. 

As conceived by Rawls the proper domain of public reason is fundamental questions about the basic structure of society in the form of constitutional essentials. These specify the roles of key societal institutions, their interrelations, and the foundational rights and liberties of citizenship. Beyond these fundamental political questions Rawls does not think public reasons are required (though it is still desirable to justify power in such terms if possible). Importantly, “many if not most political questions do not concern those fundamental matters” \cite[p.214]{RawlsPL}.  

The rationale for this restriction is the need for public reasons to be complete; they should, on their own, provide answers to most, if not all, of the questions to which they are put \cite{RawlsPL, RawlsPRR}. It is much easier to reach agreement on the fundamental questions of constitutional essentials and determine when they have been realised \cite[pp.228-9]{RawlsPL}. Consequently, among citizens, there exists a sufficiently rich overlapping consensus to justify answers concerning society's basic structure in public reasons terms.\footnote{Though philosophers have questioned whether public reasons are complete even on these fundamental questions. See for example: Reidy \citeyearpar{Reidy2000} or Horton \citeyearpar{Horton2003}.} Conversely, the ability to find public reasons which can be used to justify decisions on “more particular and detailed issues” \cite[p.230]{RawlsPL} is substantially more difficult. 

This is a problem for the legitimacy of ADS because making decisions about their design, deployment, and use requires us to address particular and detailed questions such as what constitutes an equitable outcome distribution, acceptable levels of bias and risk, and appropriate monitoring and reporting procedures \cite{Mittelstadt2019principles, Laux2023}. Answers to these questions are context specific and involve an examination of complex social, political, and economic facts about which reasonable people can disagree. Therefore, answering these questions necessitates hard choices grounded in private reasons. 

Consider, for example, recidivism prediction in which we are faced with the question of how to trade-off false positive rates against false negative rates. Prioritising false negative rates favours avoiding the incarceration of individuals who might not reoffend. Conversely, prioritising false positives focuses on community safety. We should expect reasonable people to differ in their answers to this question as answering it involves assessing various contentious social and economic facts including the impact of incarceration on employment, mental health, and interpersonal relationships, or the economic costs to the community of wrongfully imprisoning a member of society. The complexities inherent in accurately assessing and applying these facts mean that even reasonable people can arrive at different conclusions about the appropriate balance to be struck. Evaluative uncertainties which give rise to reasonable disagreement can arise for any of the decisions we need to make regarding design, deployment, and use. This suggests that, in the context of ADS, public reasons will be incomplete on a range of questions to which it is applied. 

There are two potential exceptions to such incompleteness. First, as Iason Gabriel \citeyearpar{Gabriel2022} has argued, some sociotechnical systems have become part of the basic structure of society as they influence major social institutions and practices. In these cases, we will have to address questions about the basic structure of society and public reasons will be sufficiently complete for this purpose. Secondly, completeness may not be an issue where ADS are deployed on a relatively homogenous group of decision subjects for whom it is easier to find a sufficient set of public reasons. For instance, in certain professional environments, there may be a larger set of overlapping beliefs, which we can appeal to as public reasons in the resolution of disagreement. Nevertheless, there will be numerous cases where (1) the issues to be addressed are not fundamental political questions about the basic structure of society or (2) the decision subjects affected by an ADS will not be homogeneous enough to allow for a sufficiently broad set of public reasons, for example in ADS deployed across heterogenous decision subjects in healthcare, education, credit scoring, or criminal justice settings.

The core of the challenge for public reasons as a legitimacy mechanism in ADS is the degree of abstractness and generality that is required for justifications. In practice public reasons must pass a very high bar: no reasonable person would, in theory, reject them. To some extent this difficulty is unsurprising. Theorists have consistently warned against general and abstract approaches when addressing ethical issues raised by sociotechnical systems \cite{Selbst2019, Fazelpour2020, Birhane2022, Fazelpour2022, Mittelstadt2019principles, Laux2023, young2022}.  In fact, such approaches can forestall the hard work which is required to resolve the complex normative debates which arise from new technologies \cite{Mittelstadt2019principles, Laux2023}.  

Recognising these limitations of public reason as a legitimacy mechanism for ADS, there remains one further option which may fare better. In the following section we explore theories of legitimacy grounded in democratic authorisation. This approach offers broader application, legitimating through a principle of equal influence that is applicable even in instances of intractable disagreement.\footnote{Democratic authorisation is consistent with the continued need for public reasons to justify constitutional essentials. Thus, we are not dismissing the use of public reason entirely, but advocating for a more comprehensive and practical legitimation method for questions that fall short of such fundamental political issues.} 

\section{Democratic Authorisation}
To understand the legitimising potential of democratic authorisation it is necessary to re-evaluate the objections to power discussed above. The issue is not that power constrains individual freedoms, which is an unavoidable feature of living in a collective that must reconcile the inconsistent interests of its members; rather, the objection is to the imbalanced distribution of power that allows some to make decisions for others and thus establishes unequal relations between citizens. Democracy avoids this inequality by allowing each citizen to have equal influence over decisions on their collective governance \cite{Viehoff2014, Kolodny2014II, Kolodny2014I, Kolodny2023}. 

The proposal is not that increasing democratic authorisation will somehow remedy unfairness or injustice. The call for democratic authorisation is motivated solely by its ability to equalise inequalities of power. As other scholars have noted, attempts to shift power relations come with no guarantees as to outcomes \cite{Okidegbe2021, Simonson2020}. 

The key point here is that democratic decision-making processes facilitate equal influence. But what exactly does equal influence entail? To begin, a gloss: for parties to have equal influence, any individual’s capacity to affect the outcome of a decision-making process must be equal to that of any other individual. Of course, we cannot stop at such a gloss, especially since calls to democratise AI have been criticised as undertheorized \cite{Himmelreich2023, Seger2023},  and “window dressing” \cite{Gilman2022}. Instead, the following discussion concretizes this gloss by addressing objections that equal influence is infeasible because it requires direct democracy, involvement of nonexperts in technical decision-making, and unattainable social and economic conditions. In so doing we specify how democratic authorisation ought to be implemented and the kind of decisions over which it must be applied. While more work is required to apply this theory to particular cases, we offer sufficient detail to present democratic authorisation as a promising method for legitimating ADS.

\subsection{The governance challenge}
The crux of this objection rests on the claim that if representatives are empowered to make decisions for their constituency, they possess a greater ability to determine the outcome of democratic decision-making and hence unequal influence. Consequently, if we wish to realise equal influence we should avoid representation and employ direct democracy in which decision subjects vote directly on issues. However, directly involving the potentially millions of individuals subject to ADS in decision-making is infeasible.

In response we should question the premise that representation necessarily leads to unequal influence. Why can we not design our institutions so that they minimize power differentials between representatives and their constituents? Niko Kolodny \citeyearpar{Kolodny2023} suggests we do this by making representatives beholden to the People, who should determine who represents them, the policies they enact once elected, and possess the power to contest the decisions made by their representatives.  

This provides considerable insight into the requirements for equal influence within representative democracy. First, empowering people to choose representatives necessitates free and fair elections. Furthermore, elections should be more than mere referendums on candidates; they should consider the policies candidates plan to implement. This makes campaign platforms electoral mandates, thereby constraining decision-making powers once elected \cite{Kolodny2023}. In turn this requires clear policy proposals from candidates and an informed electorate with access to the information necessary to evaluate proposals \cite{Kolodny2023}.  Lastly, to allow for the contestation of decisions, accountability mechanisms such as regular elections and appeal processes are necessary.

These requirements are substantial and are not easily realised. Yet, they are familiar to modern democracies and numerous states already have institutions that approximate or meet them. As a result, we support the suggestions of Himmelreich \citeyearpar{Himmelreich2023} and Cuéllar and Huq \citeyearpar{Huq2020} to leverage existing regulatory and legislative institutions for the democratisation of ADS. 

\subsection{The objection from expertise}
A second objection arises from Landemore and Ferreras's 2016 paper on democratising corporate governance \cite{Landemore2016}. They note scepticism towards democracy in this context because of doubts about workers' ability to govern accurately or efficiently. Applied to ADS, this argument raises similar concerns about the public (or their representatives) deciding technical questions, as they lack the specialised knowledge needed for efficient and accurate decision-making on such issues. Instead, such questions are better entrusted to experts. 

The obvious response to this objection is that it misunderstands the problem of legitimacy and returns debate to issues of expertise, efficiency, and accuracy. As we saw in Section 3, these values cannot ground legitimacy as they do not ensure grants of power are consistent with the freedom or equality of citizens. In this light, the inefficiency or inaccuracy of democratic authorisation is simply the cost we must pay for legitimate power. 

While this response is theoretically sound it does not take the objection which motivates it seriously enough. If our commitment to challenging power is more than merely theoretical, we should aim for approaches that are likely to be adopted. In commercial environments normative considerations are often ignored if they are too costly or difficult to implement \cite{Mittelstadt2019principles}. Consequently, we should seek a response which is able to reconcile democratic authorisation with the need to use experts when deciding technical questions about ADS.

Such a response is available once we realise that we do not require decision subjects or their representatives to make technical decisions about the design, deployment, and use of ADS. Note that the critical decision for legitimacy concerns who should have power and the constraints on that power. This doesn't necessitate engagement in technical decision-making. Recall that many powers gain legitimacy through acts of authorisation by state institutions. For example, insurance companies gain the power to issue insurance through legislative acts of parliament and regulations set by government agencies. These laws and regulations define the accreditation and capital requirements for insurance providers, as well as impose other restrictions, including the obligation to adhere to anti-discrimination legislation. This provides insurers with a general ability to issue insurance contracts, determine premiums, and pay out claims. State institutions, in authorising this power, do not decide individual insurance premiums. Instead, they place constraints on the provision of insurance and insurers exercise their power within those constraints. 

The same strategy can be used for ADS legitimation by allowing our existing legislative and regulatory bodies to oversee these systems, while leaving technical decisions to experts. That is, we can use these institutions to provide general overarching guidelines for the design, deployment, and use of ADS while leaving the technical decisions about how to best comply with these guidelines to experts. 

\subsection{Democratic authorisation amidst inequality}
Now that we understand what equal influence requires and the decision types it should apply to, we face our final objection: equal influence is premised upon on a “broader social ideal” \cite{Himmelreich2021} in which citizens have adequate access to the social and economic resources necessary for equal participation in decision-making. For representative government, this implies free and equal involvement in deliberation and the essential resources for informed engagement. Unfortunately, this ideal is never fully realised --- even well-functioning democracies are marked by significant disparities in economic and social resources and consequent disparities in influence over democratic processes \cite{Himmelreich2023, Huq2020, Viehoff2014, Dryzek2019}. Therefore, it seems that the requirement of equal influence establishes a standard which is infeasible. 

Does this mean that attempting to legitimate ADS using democratic authorisation is pointless? Far from it. The extent to which a decision procedure fails to realise the ideal of equal influence determines how legitimate it is \cite{Viehoff2014}.  Some processes will fall so far short as to render any decisions reached illegitimate, but others will be sufficiently egalitarian to impart a degree of legitimacy. Even imperfect democratic authorisation can go some way towards legitimating ADS. Therefore, we still have good reasons to strive for the ideal even if it is never perfectly realised. 

\section{Conclusion}
We have seen that current discourse on ethical and accountable ADS focuses on public beliefs, fairness, accuracy, efficiency and expertise, yet these aspects do not address the fundamental question of why some are permitted to wield power over others and inadvertently reinforce illegitimate power. Descriptive legitimacy allows decision-makers to rely on public perceptions rather than actually comply with the normative criteria of morally permissible power. Focusing solely on fairness, accuracy, efficiency, and expertise does not address the need to reconcile power with freedom or equality. 

In search of better foundations for our legitimacy determinations we considered some leading theories from political philosophy. Our analysis has revealed that traditional theories of legitimacy, primarily state-centric, require careful consideration before being applied to ADS. Consent is often infeasible, public reasons are not well suited to answering many of the questions we face about ADS, and democratic authorisation should form part of a broader project which addresses social and economic disparities. 

Addressing these issues so we can realise legitimacy in practice is no small task. It requires careful consideration of the particular ADS in question, a determination of the method of legitimation which is most appropriate given that context, and then careful implementation of that methodology. 

In conclusion, this paper does not claim to have resolved the issue of ADS legitimacy. Instead, it aims to catalyse further discussion and research in this area. It calls for an interdisciplinary approach, where insights from political philosophy, critical studies, feminist scholarship, and commerce converge to shape a future in which ADS are not mere tools of convenience but instruments of a free and equitable society.

\begin{acks}

For helpful comments and discussion we would like to give special thanks to Seth Lazar, Sean Donahue, Johann Laux, Sandra Wachter, Daria Onitiu, and Iason Gabriel. Various versions of this paper were presented to the Trustworthiness Auditing for AI project at the Oxford Internet Institute, the OII DPhil seminar, the Machine Intelligence and Normative Theory Lab, and the Humanising Machine Intelligence Project graduate seminar. Thank you to the conveners of those meetings, and to the audiences, for engaging with our work.

This work has been supported through research funding provided by the Wellcome Trust (grant nr 223765/Z/21/Z), Sloan Foundation (grant nr G-2021-16779), Department of Health and Social Care, and Luminate Group. Their funding supports the Trustworthiness Auditing for AI project and Governance of Emerging Technologies research programme at the Oxford Internet Institute, University of Oxford. Further support was provided by an Australian Government Research Training Program Scholarship and funds from the School of Philosophy at the Australian National University, the Machine Intelligence and Normative Theory Lab, and the Humanising Machine Intelligence Project. 

\end{acks}
\bibliographystyle{Legitimate_Power}
\bibliography{Legitimate_Power}
\end{document}